\documentclass[12pt]{iopart}

\usepackage{graphicx}
\usepackage{xfrac}
\usepackage{epsfig}
\usepackage{iopams}
\usepackage{color}
\usepackage[amssymb]{SIunits}

\begin{document}

\title[Nonlinear resonator chain]{Many Body Physics with Coupled Transmission Line Resonators}

\author{M Leib$^1$ and M J Hartmann$^1$}
\address{1 Technische Universit{\"a}t M{\"u}nchen, Physik Department, James-Franck-Str., D-85748 Garching, Germany}
\eads{\mailto{MartinLeib@circuitqed.net}, \mailto{mh@tum.de}}

\date{\today}

\begin{abstract}
We present the Josephson junction intersected superconducting transmission line resonator. In contrast to the Josephson parametric amplifier, Josephson bifurcation amplifier and Josephson parametric converter we consider the regime of few microwave photons. We review the derivation of eigenmode frequencies and zero point fluctuations  of the nonlinear transmission line resonator and the derivation of the eigenmode Kerr nonlinearities.  Remarkably these nonlinearities can reach values comparable to Transmon qubits rendering the device ideal for accessing the strongly correlated regime. This is particularly interesting for investigation of quantum many-body dynamics of interacting particles under the influence of drive and dissipation. We provide current profiles for the device modes and investigate the coupling between resonators in a network of nonlinear transmission line resonators.
\end{abstract}

\pacs{85.25.Cp, 42.50Pq, 05.30.Jp}

\maketitle 

\tableofcontents
\newpage

\section{Introduction}

The theory of interacting quantum many-body systems is a vibrant discipline in physics full of unsolved riddles like high temperature superconductivity and promising technological prospects such as topologically protected quantum states, novel schemes for quantum error correction or one way quantum computing. Minimal Hamiltonians are considered for both situations: to explain quantitatively phenomena in nature or to provide prescriptions how to build artificial systems for technological applications. 

There exists a constantly growing amount of analytical and numerical tools to investigate the properties of many body Hamiltonians. However the analytical tools typically provide answers only in limited regimes while the numeric approaches are constrained by the exponential growth of the Hilbert space with the number of particles. Dating back to an idea of Feynman \cite{feynman} who proposed to use an computer that uses the principles of quantum mechanics itself to investigate quantum mechanical many-body Hamiltonians the idea of quantum simulation \cite{quantSim} is gaining a lot of attention nowadays.  A quantum simulator is a well controllable quantum systems that emulates the physics of other systems which are less amenable to experimental investigation. Therefore quantum simulators are typically strongly enlarged versions of the ``real world'' system we try to investigate. This provides us with the possibility of tuning parameters of the Hamiltonian and with individual addressability for measurements because of larger lattice constants. Quantum many-body Hamiltonians can for example be simulated with cold atoms trapped by laser fields in various shapes and dimensions \cite{review_Bloch_2008}, in ion traps \cite{Friedenauer:2008fk,Kim2010} or arrays of cavity quantum electrodynamics (QED) systems \cite{Hartmann:2006kx,hartmann-2008-2}.

Massless photons in cavities don't interact. However by strongly coupling the photons to nonlinear systems, joint excitations emerge, called polaritons  \cite{Hartmann:2006kx,hartmann-2008-2,1367-2630-10-3-033011,PhysRevA.81.021806,Hartmann2010}, and the nonlinearity of the onsite spectrum for the polaritons can be interpreted as an interaction \cite{brandao,leib}.
Cavities can be coupled in arbitrary topologies thereby generating networks for moving photons. The polaritons inherit this abiltiy from the photons. The description of the fundamental excitations of the quantum simulator in terms of polaritons is only valid in the strong coupling regime where the coupling between the cavity and the nonlinear system exceeds the decay rates of both. In circuit quantum electrodynamics (circuit QED), the microwave realization of cavity quantum electrodynamics, one can easily achieve the strong coupling regime.
In circuit QED superconducting wires (transmission line resonators) that confine microwave photons are coupled to Josephson junction based lumped element circuits (Josephson qubits) \cite{PhysRevA.69.062320,Wallraff:2004rz,Deppe:2008kx}. Strong zero point fluctuations of the field due to the restricted dimensionality of the transmission line resonator and the artificial coupling to the Josephson qubit result in exceptionally high coupling rates \cite{Niemczyk} while the superconducting gap ensures low dissipation. 
The polariton approach has two major flaws. Firstly, the ratio of nonlinearity to hopping strength is changed by changing the polariton from purely photonic to Josephson qubit excitation. Therefore nonlinearity and hopping are mutually exclusive. Secondly, we always have two polariton species separated in energy space by the coupling strength which provides us with unwanted but unavoidable crosstalk. 

In this work, we propose making the resonator itself nonlinear by intersecting it with a Josephson junction \cite{Mallet:2009fk,leib2,Bourassa}. We show that the modes of such a device can achieve amounts of nonlinearity comparable to transmon qubits \cite{koch:042319,Majer:2007sf} for realistic circuit parameters and that the mode separation in frequency space is sufficient to suppress the crosstalk that is introduced by the nonlinearity. Moreover the coupling of these resonators is independent of the nonlinearity.  Notably, a set of experimental analysis methods for the quantum statistics of resonator networks have been demonstrated in recent years \cite{Menzel,LangEichler,Devoret}. These properties make the nonlinear transmission line resonator an ideal building block for a Bose Hubbard quantum simulator.

The article is organized as follows. First we review the derivation of frequencies and zero point fluctuations of the eigenmodes of the nonlinear transmission line resonator \cite{leib2}. Then we derive the individual Kerr nonlinearities for the eigenmodes in a low energy and rotating wave approximation. Finally we provide the current profiles including current through the inductive part of the Josephson junction and current through the transmission line resonator and discuss capacitive coupling of the eigenmodes between adjacent transmission line resonators.

\section{Nonlinear transmission line resonator}
We consider a transmission line resonator intersected in the middle by a Josephson junction.
In this section we briefly review the decomposition of the microscopic Lagrangian of the nonlinear transmission line resonator into independent modes, we provide some physical intuition for the modes and calculate their coupling strength in a network of nonlinear transmission line resonators.
\subsection{spectrum}
\begin{figure}
\includegraphics{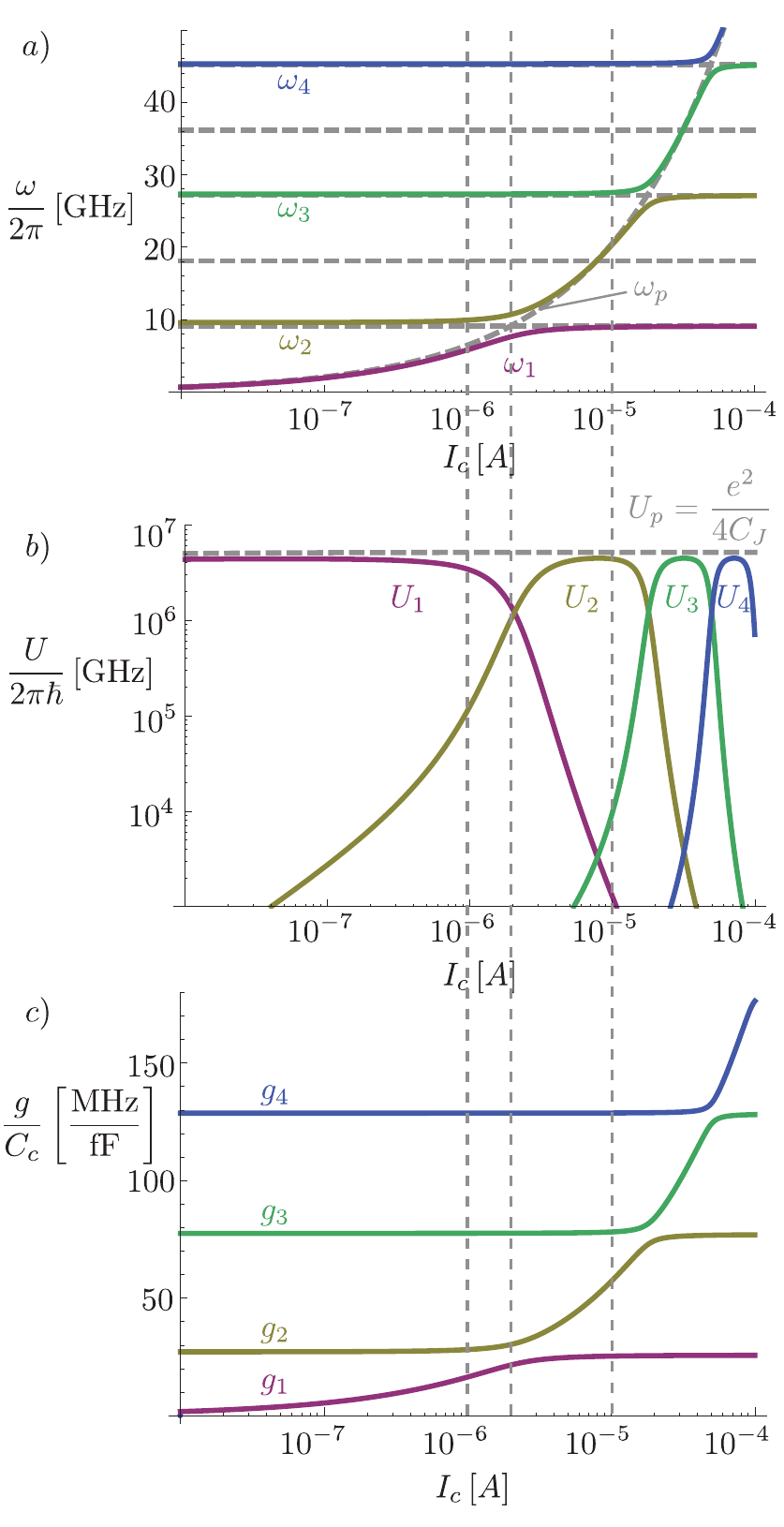}
\caption{a) Numerically calculated frequencies of the linear part of the transmission line resonator intersected by a Josephson junction. Anti-crossings show up where the plasma frequency of the Josephson junction $\omega_p$ (dashed ascending graph) matches one of the frequencies of the odd modes of the transmission line resonator (dashed horizontal graphs). b) Nonlinearity parameter of the different modes $U_n$. If the frequency of the mode is near the plasma frequency of the Josephson junction, the nonlinearity $e^2/(4 C_s)$ is inherited. c) Coupling between different antisymmetric modes of adjacent nonlinear transmission line resonators. Comparison with the mode frequencies reveals the connection of coupling strength $g_n=(C_c\omega_n)/(4\eta_n)$ and frequency of the respective mode. Dashed vertical lines mark the values of the critical current $I_c$ where we computed the current profile of the mode cf. \fref{fig:Current}}
\label{fig:specNonlin}
\end{figure}

The microscopic Lagrangian of the whole setup reads,
\begin{equation}
\mathcal{L}=\mathcal{L}_l+\mathcal{L}_r+\mathcal{L}_{JJ}\,,
\end{equation}
with the Lagrangian of the transmission line to the left and the right of the Josephson junction,
\begin{equation}
\mathcal{L}_{l/r}=\int\limits_{-\frac{L}{2}/0}^{0/\frac{L}{2}} \frac{c}{2}(\dot{\phi}_{l/r})^2-\frac{1}{2l}(\partial_x\phi_{l/r})^2 dx\,,
\end{equation}
with the flux field $\phi(x)=\int_{-\infty}^tV(x,t')dt'$ \cite{fluctuation} and the capacitance and inductance per unit length of transmission line resonator $c$ and $l$. The Lagrangian of the Josephson junction $\mathcal{L}_{JJ}$ depends on the flux drop at the Josephson junction $\delta\phi=\phi_l|_{x=0}-\phi_r|_{x=0}$,
\begin{equation}
\mathcal{L}_{JJ}=\frac{C_J}{2}\delta\dot{\phi}^2+E_J\cos(\frac{2\pi}{\phi_0}\delta\phi)\,,
\end{equation}
where $C_J$ and $E_J$ are the capacitance and Josephson energy of the junction respectively. $\phi_0=h/2e$ is the quantum of flux. We proceed by separating the linear from the purely nonlinear parts of the Lagrangian. This way we can exactly diagonalize the linear parts and get the normal modes of the system. We then reintroduce the nonlinear parts as an perturbation to the oscillator modes of the system. It suffices to consider the nonlinear cosine inductance of the Josephson junction when separating the nonlinear parts from the linear ones ,
\begin{equation}
E_J\cos(\frac{2\pi}{\phi_0}\delta\phi)=-\frac{1}{2L_J}\delta\phi^2+E_J\left(1+\sum\limits_{n=2}\frac{-1^n}{(2n)!}(\frac{2\pi}{\phi_0}\delta\phi)^{2n}\right)\,,
\end{equation}
where $L_J=\phi_0^2/(4\pi^2E_J)$ is the Josephson inductance.
Considering only the linear part of the Lagrangian, the Euler Lagrange equations provide us with the wave equation in the two arms of the transmission line resonator,
\begin{equation}
\partial_t^2\phi_{l/r}=v^2\partial_x^2\phi_{l/r}\,,
\end{equation}
where the phase velocity is given by the inductance and capacitance per unit length $v=1/\sqrt{cl}$. Boundary conditions for the flux field can either be derived by Kirchhoffs rules or also via the Euler Lagrange equations.  The current in the transmission line resonator is proportional to the spatial derivative of the flux field $I=\frac{1}{l}\partial_x\phi$ and  has to vanish at the ends of the transmission line,
\begin{equation}
\partial_x \phi_{l/r}|_{x=\mp\frac{L}{2}}=0.
\end{equation}
We also know that the current flowing into the Josephson junction from one side has to exit on the opposite side,
\begin{equation}
\partial_x\phi_l|_{x=0}=\partial_x\phi_r|_{x=0}\,.
\end{equation}
The current flowing through the junction can either flow into the shunting capacitor or the Josephson junction. The current flowing through the junction is related to the flux drop across the junction in the Josephson constitutive relation $I_{J}=I_c\sin((2\pi/\phi_0)\delta\phi)$. The current into the capacitor in turn reads, $I_{cap}=C_J\delta\ddot{\phi}$. For the linearized Lagrangian the condition for the current flowing through the Josephson junction reads,
\begin{equation}
-\frac{1}{l}\partial_x \phi_l=C_J\delta\ddot{\phi}+\frac{1}{L_J}\delta\phi\,.
\end{equation} 
To find the spatial eigenmodes of the nonlinear transmission line resonator we first make an ansatz with separation of variables, $\phi(x,t)=f(x)g(t)$ with $f_l(x)=A_l\cos(k(x+L/2))$ for the transmission line to the left of the Josephson junction and $f_r(x)=A_r\cos(k(x-L/2))$ for the right part of the transmission line. With this ansatz the necessity of vanishing currents at the ends of the transmission line is automatically satisfied and the wave equation is fulfilled provided that $\omega=k v$, where $\omega$ is the frequency of the temporal part of the flux function $\ddot{g}+\omega^2 g=0$. There are two distinct possibilities to fulfill the boundary conditions at the Josephson junction: we could either choose $A_l=A_r$ which provides us with the symmetric modes of the resonator or we could choose $A_l=-A_r$ to get the antisymmetric modes. The symmetric modes show no flux drop at the Josephson junction and consequently there is no current flowing through the junction. Therefore the symmetric modes of the transmission line resonator are completely unaffected by the presence of the Josephson junction and we omit them in the further discussion of the system dynamics. For the antisymmetric modes we choose the amplitudes to be $A_l=-A_r=1$ and get the following transcendental equation for the mode frequencies $\omega_n=k_n v$,
\begin{equation}
\frac{\omega_n}{v}=\cot\left(\frac{\omega_n}{v}\frac{L}{2}\right)\frac{2l}{L_J}\left(1-\frac{\omega_n^2}{\omega_p^2}\right)\,,
\end{equation}
where $\omega_p=1/\sqrt{L_JC_J}$ is the Josephson plasma frequency. The solution of the above transcendental equation is the spectrum of the system, not taking into account the symmetric modes, presented in \fref{fig:specNonlin} a) for generic parameters given in \tref{tab:ExpParam}. We plotted the spectrum as a function of the critical current $I_c$ of the Josephson junction. The resonator modes without the Josephson junction do not depend on the cricital current of the Josephson junction but the plasma frequency $\omega_p=\sqrt{I_c/(\phi_0 C_J)}$ does. If the plasma frequency and the frequency of the antisymmetric modes of the bare transmission line resonator (without Josephson junction) are detuned the frequencies of the combined system are the frequencies of its constituents, Josephson junction or transmission line resonator. But if the plasma frequency is resonant with one of the symmetric modes we get an anticrossing of the two frequencies which is of the order of the mode frequency itself accounting for the large coupling between Josephson junction and transmission line resonator. 

\begin{table}
\caption{\label{tab:ExpParam} }
\begin{indented}
\item[] \begin{tabular}{@{}rll}
\br
\multicolumn{3}{l}{transmission line} \\
\mr
\hspace{10mm} characteristic impedance & $ Z_0 $ & $\unit{50}{\ohm}$\\
phase velocity & $v$ & $\unit{0.94 \cdot 10^8}{\meter\per\second}$\\
inductance per length & $l$ & $\unit{5\cdot 10^{-7}}{\henry\per\meter}$\\
capacitance per length & $c$ & $\unit{2 \cdot 10^{-10}}{\farad\per\meter}$\\
\br
\multicolumn{3}{l}{Josephson Junction}\\
\mr
shunting capacitance&$C_J$& $\unit{1.9 \cdot 10^{-12}}{\farad}$\\
\br
\end{tabular}
\end{indented}
\end{table}

We deduce the following generalized ``scalar  product'' between the spatial modes with the help of the transcendental equation of the mode frequencies,
\begin{equation}
c\int_{-\frac{L}{2}}^{\frac{L}{2}}f_n(x)f_m(x)dx+C_s\delta f_n \delta f_m = \delta_{n,m} \eta_n\,,
\end{equation}  
with,
\begin{equation}
\eta_n=c\left(\frac{L}{2}+\frac{\delta f_n^2}{k_n^2}\frac{l}{2L_J}\left(1+\frac{\omega_n^2}{\omega_p^2}\right)\right)\,,
\end{equation}
where $\delta f_n = f_{n,l}|_{x=0}-f_{n,r}|_{x=0}$ is the normalized flux drop of the spatial modes. We use this ``scalar product'' to decompose the linearized Lagrangian into independent oscillators of frequencies $\omega_n$ and masses $\eta_n$. After a Legendre transformation and omitting constant terms we get the Hamiltonian,
\begin{eqnarray*}
H&=&\sum_{n=1}^{\infty}\left(\frac{\pi_n^2}{2\eta_n}+\frac{1}{2}\eta_n \omega_n^2g_n^2\right)-E_J\sum\limits_{n=2}\frac{-1^n}{(2n)!}(\frac{2\pi}{\phi_0}\delta\phi)^{2n}\\
\pi_n&=&\eta_n\dot{g}_n\,. 
\end{eqnarray*}
Up to here everything we did was classical physics, now we proceed by quantizing the Hamiltonian in the usual way by introducing independent raising and lowering operators for the different modes,
\begin{eqnarray*}
\hat{\pi}_n&=&-i\sqrt{\frac{\eta_n\omega_n}{2}}(a_n-a_n^{\dag})\\
\hat{g}_n&=&\frac{1}{\sqrt{2\eta_n\omega_n}}(a_n+a_n^{\dag})\,,
\end{eqnarray*}
with $\left[a_n,a_m^{\dag}\right]=\delta_{n,m}$. Finally, we arrive at the following Hamilton operator,
\begin{eqnarray}
H&=&\sum_{n=1}^{\infty}\omega_n(a_n^{\dag}a_n+\frac{1}{2})+H^{\text{nonlin}}\nonumber\\
H^{\text{nonlin}}&=&-E_J\left(\cos\left(\delta\tilde{\phi}\right)+\frac{1}{2}\delta\tilde{\phi}^2\right)\,,\label{eq:unHarmHam}
\end{eqnarray}
with 
\begin{equation*}
\delta\tilde{\phi}=\sum_{n=1}^{\infty}\lambda_n\left(a_n+a_n^{\dag}\right)\quad \text{and,} \quad
\lambda_n=\frac{2\pi\delta f_n}{\Phi_0\sqrt{2\eta_n \omega_n}}\,.
\end{equation*}
The nonlinearity again reintroduces interactions between the modes and nonlinearities for the individual modes. As we will show below these are all much smaller than the differences between the mode frequencies. They only play a role in strong driving scenarios like the Josephson parametric converter. We are only interested in few microwave photon physics and therefore may omit the residual coupling in a rotating wave approximation. The individual mode nonlinearities however can not be omitted. For their derivation we start to decompose the cosine term with a product-to-sum formula which is applicable in this case because all individual flux operators commute,
\begin{eqnarray*}
\cos\left(\sum\limits_{n=1}^{\infty} \lambda_n(a_n+a_n^{\dag})\right)&=&\cos\left(\lambda_1(a_1+a_1^{\dag})\right)\cos\left(\sum\limits_{n=2}^{\infty} \lambda_n(a_n+a_n^{\dag})\right)\\
&&-\sin\left(\lambda_1(a_1+a_1^{\dag})\right)\sin\left(\sum\limits_{n=2}^{\infty} \lambda_n(a_n+a_n^{\dag})\right)\,.
\end{eqnarray*} 
The sine terms only contain rotating operator products and therefore can be neglected in the framework of the rotating wave approximation. If we further proceed this way we may transform the cosine of the sum of all mode flux operators into the product of the cosines of all mode flux operators,
\begin{equation*}
\cos\left(\sum_n \lambda_n (a_n+a_n^{\dag})\right)\to \prod_n \cos\left(\lambda_n(a_n+a_n^{\dag})\right)\,.
\end{equation*}
The cosines of the mode flux operators may be rewritten in a sum of normal ordered products of mode operators. We may further also omit all operator products containing a different amount of raising and lowering operators, again because of the rotating wave approximation. Finally we get as a rotating wave, low energy approximation for the cosine of a mode flux operator,
\begin{equation*}
\cos\left(\lambda_n(a_n+a_n^{\dag})\right)\to e^{-\frac{\lambda_n^2}{2}}\left(1-\lambda_n^2 a_n^{\dag}a_n+\frac{\lambda_n^4}{4} a_n^{\dag} a_n^{\dag}a_na_n+\dots\right)\,.
\end{equation*}
We concentrate on the fundamental mode and consider all the other modes to be in the vacuum state which leads us to an Hamiltonian only describing the fundamental antisymmetric mode of the transmission line resonator coupled to the plasma mode of the Josephson junction,
\begin{equation*}
H_1=\left(\omega_1-\delta \omega\right) a_1^{\dag}a_1-U a_1^{\dag}a_1^{\dag}a_1a_1\,,
\end{equation*}
with
\begin{eqnarray*}
\delta \omega &=& E_J \lambda_1^2\left(1-\prod\limits_{n=1}^{\infty} e^{-\frac{\lambda_n^2}{2}}\right)\\
U &=& E_J \frac{\lambda_1^4}{4}\prod\limits_{n=1}^{\infty} e^{-\frac{\lambda_n^2}{2}}\,,
\end{eqnarray*}
where $\delta\omega$ is a small renormalization  of the fundamental mode frequency due to the nonlinearity. The Kerr nonlinearity $U$ is plotted in \fref{fig:specNonlin} b) as a function of the Josephson junctions critical current $I_c$. Direct comparison with the spectrum of the system reveals the origin of the nonlinearity. For values of $I_c$ where the respective frequency is defined by the plasma frequency of the Josephson junction, the mode inherits the full nonlinearity of the Josephson junction.
\subsection{modes}

\begin{figure}
\includegraphics{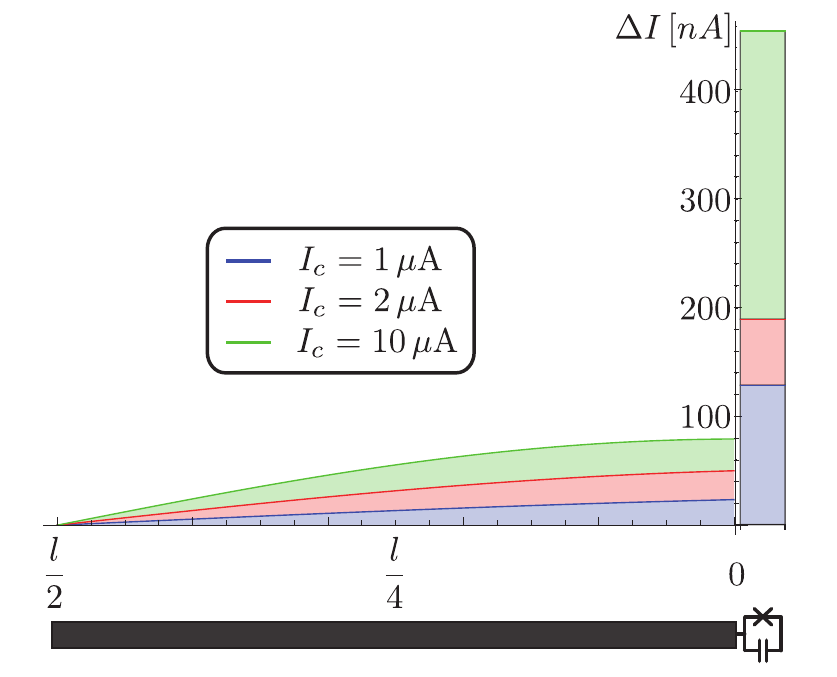}
\caption{Variance of the current in the transmission line resonator and Josephson junction for the first Fock state of the fundamental mode.}
\label{fig:Current}
\end{figure}

We now decomposed the linear part of the Lagrangian into independent harmonic oscillators with frequencies $\omega_n$ and mode capacitances $\eta_n$. The latter only make sense in combination with the normalization convention we chose for the spatial mode functions $f_n(x)$. In order to get the correct physical intuition of the oscillation mode, we have to revert to the well known quantities of oscillating circuits, currents and charges or rather coulomb potential. We start by calculating the current through the Josephson junction. The observable for the current through the Josephson junction is,
\begin{equation}
\hat{I}_J=I_c\sin(\delta\tilde{\Phi})\,.
\end{equation}
Lets suppose that all modes except for the fundamental mode are in their vacuum state and the fundamental mode is in a Fock state $\left|n \,0\dots 0\right\rangle$. In this state the mean value of current flowing through the Josephson junction vanishes comparable to the vanishing displacement of a quantum harmonic oscillator in a Fock state. The variance of the current however doesn't vanish and provides us with an estimation of the mean current flowing through the Josephson junction,
\begin{equation}
\Delta I_J=I_c\sqrt{\left\langle n\,0\dots0\right|\sin(\delta\tilde{\phi})^2\left|n\,0\dots 0\right\rangle}\,.
\end{equation}
With the same arguments as presented above for the derivation of the Kerr parameter, we calculate the variance, which is exact up to the second Fock state,
\begin{equation}
\Delta I_J=I_c \sqrt{\frac{1}{2}\left(1-\prod_k e^{-2\lambda_k^2}\left(1-4\lambda_1^2 n + 4\lambda_1^4 n (n-1)\right)\right)}\,.
\end{equation}
The observable for the current in the transmission line to the left and the right of the Josephson junction is,
\begin{equation}
\hat{I}_r(x)=\frac{1}{l}\partial_x \hat{\phi}(x)=\sum_n\frac{\partial_x f_n(x)}{l\sqrt{2 \eta_n \omega_n}} (a_n+a_n^{\dag})\,.
\end{equation}
Here again the mean current in the transmission line resonator for a Fock state vanishes and we evaluate the variance of the current, 
\begin{equation}
\Delta I_r=\frac{1}{l}\sqrt{ \frac{\left|\partial_x f_1(x)\right|^2}{2 \eta_1 \omega_1}( 2n + 1)+\sum\limits_{k=2}^{n_{cutoff}}\frac{\left|\partial_x f_k(x)\right|^2}{2 \eta_k \omega_k}}\,.
\end{equation}
\fref{fig:Current} shows the variance of the current in the left half of the nonlinear transmission line resonator and the inductive part of the Josephson junction. The variance of the current in the right part is for symmetry reasons the mirror image of its counterpart in the left half.  
\subsection{coupling}
Next we are calculate the strength of the coupling between two neighboring transmission line resonators in a network.  We only consider capacitive coupling of resonators where the respective ends of the central lines of adjacent transmission line resonators are either close to each other or connected by interdigitated capacitors. To integrate the coupling into our theoretical model we have to include the energy of the coupling capacitor, with coupling capacitance $C_c$, into the Lagrangian of the nonlinear transmission line resonator,
\begin{equation}
\mathcal{L}_c=\frac{C_c}{2}\left(\dot{\phi}_1|_{x=\frac{L}{2}}-\dot{\phi}_2|_{x=-\frac{L}{2}}\right)^2\,,
\end{equation}
where $\phi_{1/2}$ is the flux field of adjacent nonlinear transmission line resonators. After a Legendre transformation to get the corresponding energy term in the Hamiltonian we would get different conjugate momenta $\pi_n$. We neglect this effect because the coupling capacitance $C_c$ is very small compared to the overall capacitance of the nonlinear transmission line resonator and get $\dot{\phi}_{1/2}=-i\sum_n\sqrt{\omega_n/(2\eta_n)}f_n(x)\left(a_{1/2,n}-a_{1/2,n}^{\dag}\right)$. The coupling term in the Lagrangian therefore provides us with many different effects: We get renormalizations of the different resonator mode frequencies, exchange couplings between different modes of adjacent resonators and on the same resonator, which we neglect in a rotating wave approximation, and exchange coupling of the same mode of adjacent resonators,
\begin{equation}
H_{n,c}=-\frac{C_c}{4}\frac{\omega_n}{\eta_n}\left(a_{1,n}^{\dag}a_{2,n}+a_{1,n}a_{2,n}^{\dag}\right)\,.
\end{equation}
The dimensionless capacitance of the fundamental modes of the nonlinear transmission line resonator is not affected by a change of the Josephson junction's critical current for generic parameters (cf. \tref{tab:ExpParam}) of our setup . The coupling $g_n=(C_c \omega_n)/(4 \eta_n)$ of the same modes in adjacent resonators is therefore determined by the frequency of the modes cf. \fref{fig:specNonlin} c), where we plotted the coupling of the first three antisymmetric modes of the nonlinear transmission line resonator. This in turn enables us to increase the nonlinearity while increasing the coupling, provided we do not use the fundamental mode of the nonlinear transmission line resonator. 

\section{Summary}
We reviewed the derivation of the eigenmode frequencies and respective zero point fluctuations for the Josephson junction intersected transmission line resonator. We derived the Kerr nonlinearities of the individual eigenmodes and calculated  the current in the transmission line resonator and the Josephson junction. Finally we showed that the coupling depends on the eigemode frequency and is thereby independent of the nonlinearity. This lack of mutual exclusivity of nonlinearity and coupling makes quantum simulators with nonlinear transmission line resonators superior to polariton approaches.

\noappendix

\end{document}